\documentclass[singlecolumn,showpacs,amsmath,amssymb,article,superscriptaddress]{revtex4}
\usepackage{epsfig}
\usepackage{amsmath}
\usepackage{amssymb}
\usepackage{graphicx}
\usepackage{dcolumn}
\usepackage{bm}

\def\be{\begin{eqnarray}}
\def\ee{\end{eqnarray}}
\def\ba{\begin{array}}
\def\ea{\end{array}}

\begin{document}
\title{The local nature of incompressibility at quantised Hall effect modified by interactions}
\author{E. M. Kendirlik}
\affiliation{Department of Physics, Faculty of Sciences, Istanbul University, 34134-Vezneciler, Istanbul, Turkey}
\author{S. Sirt}
\affiliation{Department of Physics, Faculty of Sciences, Istanbul University, 34134-Vezneciler, Istanbul, Turkey}
\author{S. B. Kalkan}
\affiliation{Department of Physics, Faculty of Sciences, Istanbul University, 34134-Vezneciler, Istanbul, Turkey}
\author{N. Ofek}
\affiliation{Applied Physics Department, Yale University, 217 Prospect Street
New Haven, CT 06511-8499, USA}
\author{V. Umansky}
\affiliation{Braun Center for Submicron Research, Department of Condensed Matter Physics, Weizmann Institute of Science, Rehovot 76100, Israel}
\author{A. Siddiki}
\affiliation{Department of Physics, Faculty of Sciences, Istanbul University, 34134-Vezneciler, Istanbul, Turkey}

\begin{abstract}
Since the experimental realisation of the integer quantised Hall effect in a two dimensional electron system subject to strong perpendicular magnetic fields in 1980, a central question has been the interrelation between the conductance quantisation and the topological properties of the system. It is conjectured that if the electron system is described by a Bloch hamiltonian, then the system is insulating in the bulk of the sample throughout the quantised Hall plateau due to magnetic field induced energy gap. Meanwhile, the system is conducting at the edges resembling a 2+1 dimensional topological insulator without the time-reversal symmetry. However, the validity of this conjecture remains unclear for finite size, non-periodically bounded real Hall bar devices. Here we show experimentally that the close relationship proposed between the quantised Hall effect and the topological bulk insulator is prone to break for specific magnetic field intervals within the plateau evidenced by our magneto-transport measurements performed on GaAs/AlGaAs high purity Hall bars with two inner contacts embedded to bulk. Our data presents a similar behaviour also for fractional states, in particular for 2/3, 3/5 and 4/3.

\end{abstract}

\maketitle

The mutual relation between the integer quantised Hall effect~\cite{vKlitzing80:494} (IQHE) and topology is a reoccurring theme.~\cite{Thouless82:405,Dassarma,Girvin,Hasan:10:3045} Salient features of the IQHE are the precise Hall conductance measured as integer multiples of the conductance quanta $e^2/h$ ($e$ is the elementary charge and $h$ is the Planck constant) accompanied by zero longitudinal resistance at certain magnetic field intervals. The robustness of the IQHE against material systems points to a universal origin, which is claimed to be the topology of the system~\cite{Dassarma} and relies on a key argument: the bulk of the two dimensional electron system (2DES) is incompressible.~\cite{Girvin} For the IQHE, the incompressible state is a direct consequence of the quantising magnetic field $B$ localising the electrons when the Fermi energy $E_F$ falls a Landau gap. Hence, due to the lack of available states the bulk is insulating, while the number of Landau levels below $E_F$ determines the filling factor $\nu$. In the case of the fractional quantised Hall effect the energy gap emanates from many-body interactions, based on exchange and correlation effects, and $\nu$ assumes some particular fractional numbers.~\cite{Hasan:10:3045,FQHE} Quite generally, the bulk insulating region is assumed to be incompressible and provides a scattering free region between the probe contacts. 

The rationale behind the relation between the incompressible topological bulk insulator and the conductance quantisation stems from the mathematical map between the Landau Hamiltonian and the Bloch Hamiltonian defined on a periodic system, which can be well understood physically in terms of the Berry phase.~\cite{Hasan:10:3045} Each time the field is increased by one magnetic flux quantum $\Phi_0$ $(=e/h)$, an electron is adiabatically transferred from one edge of the cylinder to the other edge (cf. Supplementary Material Fig.4), keeping the geometrical phase protected. In the standard picture, the charge transport along the edges is described by the single particle, i.e., non-interacting, edge states which form due to level bending imposed by the boundary conditions. The transport is regarded as non-local and dissipationless.~\cite{Buettiker86:1761} Assuming periodic boundaries, the Hall conductance $\sigma_{xy}$ is a topological invariant, namely the Chern number,  which can be elegantly proven using the Kubo formalism. In the case of the IQHE with $\sigma_{xy}=\nu\frac{e^2}{h}$ the filling factor $\nu$ is the Chern number.~\cite{Thouless87:101} 
Within the single-particle approaches, adding two inner contacts to the bulk of the system modifies the topology. Hence, conductance quantisation should be
severely affected, unless the bulk insulating state remains unchanged.
To probe whether the bulk remains incompressible one can impose an external excitation between the inner contacts and measure the potential difference as a function of magnetic field. Different from typical Hall experiments, here one does not only measures the resistance $R$ but measures the impedance $Z$ to explore the contribution from the capacitance $C$. The impedance is evaluated as $Z=\sqrt{R^2-1/(2\pi i f C)^2}$, $f$ being the frequency of the AC excitation. Here, the (quantum) capacitance is given by $C=e^2D_{\rm T}(E_F)$, and $D_{\rm T}(E_F)$ is the thermodynamic density of states (TDOS) at $E_F$. Since, the bulk is incompressible due to $D_{\rm T}(E_F)=0$ throughout the conductance plateau, the capacitance vanishes and impedance diverges to infinity for an ideally pure 2DES at $T\rightarrow0$.

\begin{figure}[ht!]
\includegraphics[width=1.\columnwidth]{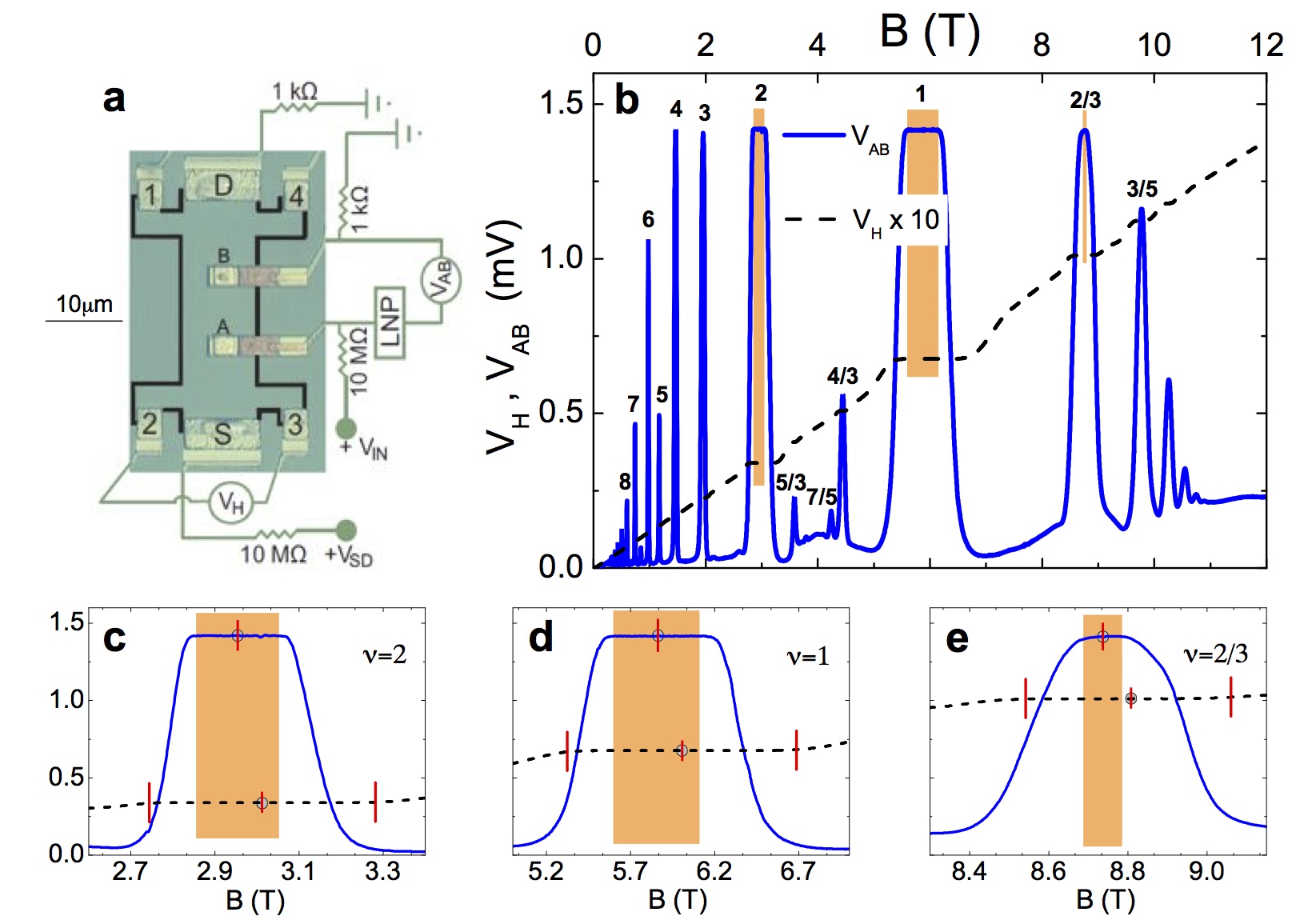}
\caption{\label{fig:1}\textbf{The experimental setup and measured voltages at base temperature.} \textbf{a,} The SEM image of the Hall bar defined on high purity GaAs/AlGaAs wafer, together with the measurement setup. A 26 mV$_{RMS}$ AC voltage excitation is imposed between source-drain contacts (S and D, at 8.54 Hz)) $V_{\rm SD}$ and inner contacts (A-B, 4 mV$_{RMS}$ at 11.5 Hz) $V_{\rm in}$, simultaneously. The Hall potential $V_{\rm H}$ is measured between contacts 1-4 or 2-3, whereas longitudinal voltage difference is measured between contacts 1-2 or 3-4. \textbf{b,} The Hall potential measured between contacts 1-4 (broken (black) line) and $V_{\rm AB}$ (thick solid (blue) line) as a function of magnetic field $B$, while imposing AC excitations between contacts A-B and S-D. We also checked that there is no correlation between $V_{\rm AB}$ and $V_{\rm SD}$ by observing the measured potential differences without one of the excitations. Shaded ares depict the constant $V_{\rm AB}$ subintervals observed for $\nu=2,1$ and 2/3, which are zoomed in \textbf{c, d, e}, respectively. The vertical long-lines indicate the plateau intervals, whereas their relative centres are indicated by vertical short-lines with circles. The constant voltage intervals are determined by a percentage error of \% 1, corresponding to an absolute error less than $10^{-8}$.}
\end{figure}

In contrast to the above single particle description of the IQHE, it is proposed that the direct Coulomb interaction modifies the electronic distribution and the bulk becomes compressible also within the plateau, i.e. the bulk behaves like a metal with high density of states at $E_F$.~\cite{Chang90:871,Chklovskii92:4026,siddiki2004} In this situation, the incompressible states (or commonly called strips) reside at the edges of the sample, due to homologous level bending of the 1D edge-states. Under these conditions, it is predicted that one can drive an excitation between the inner contacts and measure a finite impedance, without disturbing the perfect quantisation.~\cite{Aylin:unpub} Inspired by this proposal, it is interesting to determine the effects of inner contacts on the Hall conductance when imposing an excitation between them. It is possible to find few similar attempts in the literature to clarify the incompressibility of the bulk, however, most of the experiments utilise indirect measurement techniques, such as capacitive coupling, scanning force microscopy or single electron transistors,~\cite{Suddard:10,Ahlswede02:165,Yacoby04:328} where the samples usually are either low purity ($<10^6$ cm$^2$/Vs) and/or large in dimensions ($W>100 \mu$m).

Here we investigate directly the bulk transport properties of a topologically modified Hall bar using our TLM 400 dilution fridge equipped with a 20 Tesla super-conducting magnet at low temperatures ($\leq750$ mK). The potential difference between the inner contacts is measured simultaneously with a standard Hall characterisation while an AC excitation voltage is imposed. We find that the bulk of the Hall bar is not entirely insulating all along the conductance plateaus. This observation is in perfect agreement with the predictions of the interaction theory of the IQHE, however, challenges the single particle theories, which incontestably assume an insulating bulk in the plateau interval regardless of the magnetic field strength. The experimental findings will lead to reexamine the existing theories of the IQHE  
from a more general topological point of view and open new a horizon in investigating the relation between topological insulators and quantised Hall effect.

\noindent\textbf{Experiments at base temperature}

\noindent In Fig.\ref{fig:1}a we show the SEM image of our 10 $\mu$m wide Hall bar together with the sketch of the experimental setup. The 2DES is created in a high purity ($\mu\approx 8\times10^6$ cm$^2$/Vs, after illumination) GaAs/AlGaAs heterostructure residing 130 nm below the surface. The transport experiments are repeated for three different thermal cycles, at dark and after illumination considering two different Hall bars to eliminate specific sample properties. We found that all measurements present very similar salient features. Fig.\ref{fig:1}b depicts measured $V_{\rm H}$ (broken lines) between contacts 1 and 4, where a 26 mV root mean square $V_{\rm SD}$ excitation is imposed at base temperature $\lesssim15$ mK (electron temperature $\lesssim$ 80 mK), corresponding to an excitation current of 0.4 nA, sufficiently small to prevent heating effects. The thick solid line depicts $V_{\rm AB}$, where couple of spikes for $B\lesssim2$ T is observed, each corresponding to an integer $\nu$. Similar spike-like features are also observed in the intervals $3.5$ T $\lesssim B\lesssim5$ T and $8$ T $\lesssim B\lesssim11$ T for which one can assign fractional states with $\nu=$ 11/7, 4/3, 3/5 , etc. In contradistinction to mentioned spikes, for $\nu=2$, 1 and 2/3 we observe a constant $V_{\rm AB}$ (highlighted by shaded ares) in a subinterval of the corresponding Hall plateau. The related filling factors are enlarged at the bottom panel, Fig.~\ref{fig:1}(c)-(e). We attribute this behaviour to a well established bulk incompressible region as we will explain in the following. Even grippingly, on both sides of the plateaus $V_{\rm AB}$ takes values similar to that of which are completely out of the plateau intervals. A homologous behaviour is also observed for the fractional state 2/3, as shown in Fig.~\ref{fig:1}e. The variation of $V_{\rm AB}$ is of primary importance pointing a compressible bulk in spite of a simultaneously measured quantised Hall plateau. After a first glance at the experimental data, the questions are: Why is $V_{\rm AB}$ constant only for a specific $B$ subinterval and is it quantised? What is the mechanism behind the observed asymmetry with respect to the centre of the plateau and how the quantised $V_{\rm H}$ and $V_{\rm AB}$ compare with longitudinal potential difference? In the next Section we will answer these questions within the screening theory of the IQHE.

\noindent \textbf{Discussion}

\noindent We base our model on the formation of the compressible and the incompressible strips resulting from electron-electron interactions and the confinement. Since the non-interacting single particle picture yields a stepwise density distribution violating the electrostatic equilibrium (cf. Supplementary Material Fig.5), it is propounded that the 2DES comprises local insulator-like incompressible regions surrounded by metal-like compressible regions. Strips with integer $\nu$ emanate from direct Coulomb (Hartree) interaction, whereas, the strips assuming fractional filling factors account on many-body interactions. Taking into account interactions and lateral confinement results in the fact that the incompressible region(s) can reside at the bulk, as well as along the edges. The former is called the bulk incompressible region, whereas the later is called the edge incompressible strip. The spatial distribution and the widths of the incompressible strips are determined by the external $B$ field, the density gradient, the temperature and most importantly by the energy gap, as clarified both analytical and self-consistent numerical calculations.~\cite{Chklovskii92:4026,siddiki2004} Assuming a constant $E_F$, the evolution of the incompressible regions with changing $B$ field can be summarised as follows: Let us start with a situation where the field is sufficiently high and only the lowest Landau level is partially filled, hence, the average filling factor is less than 2 (neglecting the spin degeneracy). Due to the lateral confinement, the spatial distribution of the electron density is inhomogeneous such that it reaches its highest value close to the centre (at the bulk) whereas gradually decreases towards the edges and vanishes at the boundaries. Therefore, $\nu_{\rm center}$ is larger than that of at the edges. 
\begin{figure}[t!]
\includegraphics[width=1.\columnwidth]{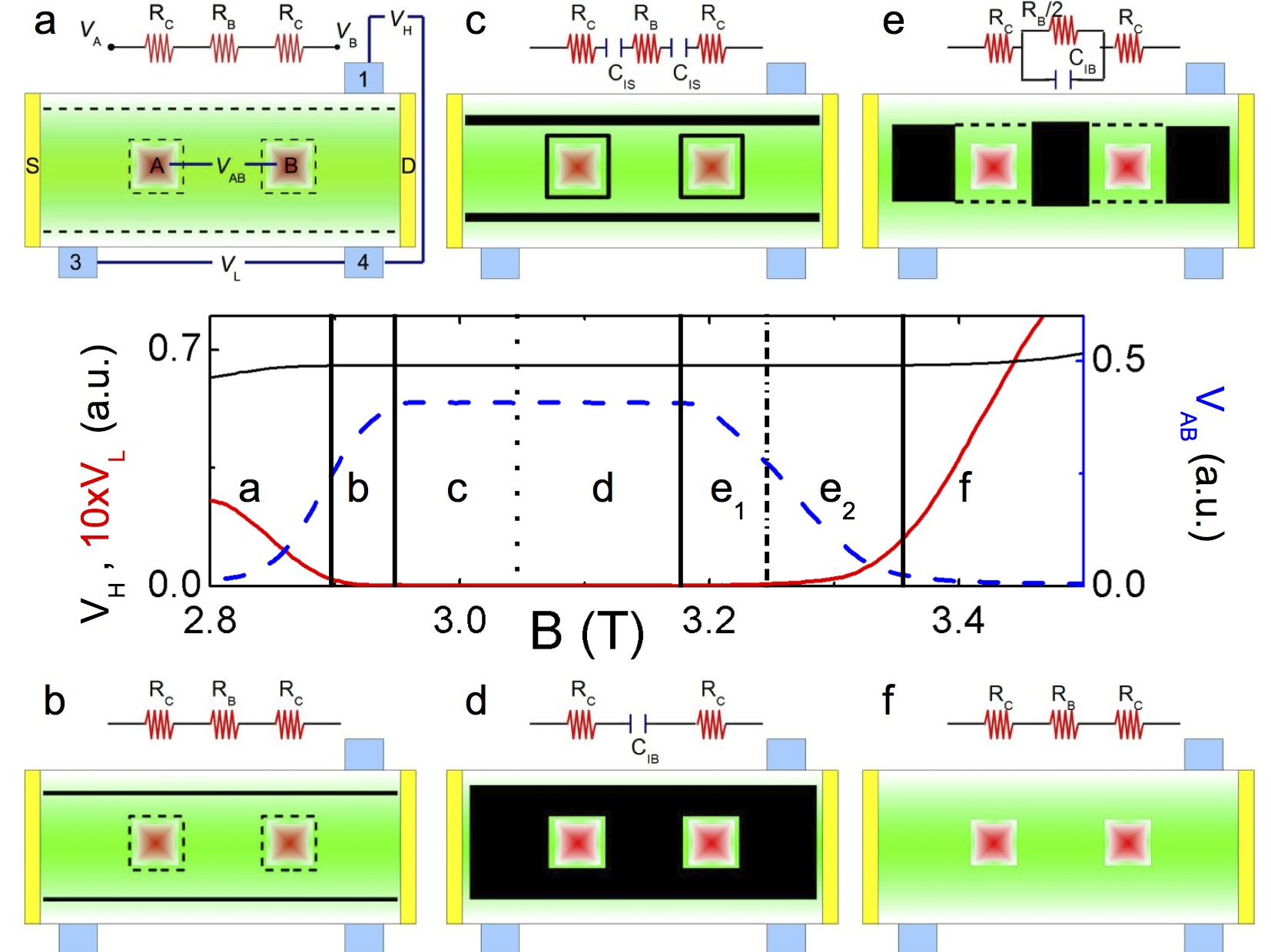}
\caption{\label{fig:2}\textbf{Graphical sketches of the electron density together with the measured voltages.} Central figure: The measured voltages $V_{\rm H}$ (solid thin (black) line), $V_{\rm L}$ (solid thick (red)  line) and $V_{\rm AB}$ (broken (blue) line). Vertical lines are guide to eye to separate different cases depicted in \textbf{a-f}. In graphical sketches, colour gradient depicts the electron density distribution and boxes denote contacts (blue: probe contacts - red: inner contacts - yellow: source and drain contacts.). Black regions identify incompressible regions, whereas dashed lines correspond to thermodynamically compressible strips. \textbf{b-f}, The spatial distribution and widths of incompressible regions together with corresponding electric circuit of each case. Here $R_C$ and $R_B$ denote the contact and compressible bulk resistances respectively, while $C_{IB}$ depicts the bulk capacitance and $C_{IS}$ the capacitance of the encircling strips.}
\end{figure}
Once the $B$ field is decreased, the Fermi energy falls into the Landau gap at the bulk yielding $\nu_{\rm centre}=2$, hence the bulk becomes incompressible. This resembles  the single-particle description of the IQHE, where the $V_{\rm H}$ is quantised and $V_{\rm L}$ vanishes. Further decrease of the field results in a situation where the incompressible regions shift towards edges and become narrow, due to the electron density gradient. The transport properties and evolution of strips along the edges are the well understood both experimentally and theoretically,\cite{Ahlswede02:165,siddiki2004} and guarantee quantised $V_{\rm H}$ together with vanishing $V_{\rm L}$ by the thermodynamical and the electrical decoupling of the probe contacts. Namely, scattering between contacts is prohibited by the incompressible regions between them. Here we used the fundamental principle of thermodynamics in determining the (in)compressibility of the strips: Thermodynamical quantities are physically meaningful only if there are sufficiently large number of particles, which dictates a lower bound for the length scales. At $T\rightarrow0$ this length is the Fermi wavelength, $\lambda_F$, since only the electrons at $E_F$ contribute to current. Therefore once the width of the incompressible strip becomes comparable or less than $\lambda_F$, assigning incompressibility to strips become meaningless, physically. Hence, it is thermodynamically reasonable to expect that at a certain $B$ field the width of incompressible strips become narrower than $\lambda_F$ and scattering between neighbouring compressible regions and thereby the probe contacts is possible. In this situation, $V_{\rm L}$ becomes finite and $V_{\rm H}$ deviates from its quantised value. The schematic presentation of the above discussion is depicted in Fig.\ref{fig:2}, with increasing $B$ from (a) to (f) where the filling factor distribution of a Hall bar is shown which comprises two inner contacts. The electron density gradient is depicted by the colour gradient, whereas the incompressible regions are shown by black. Similar to edges, incompressible strips form in the close vicinity of inner contacts, where the broken lines denote the case where the strips become `leaky" both thermodynamically and electrically. 
\begin{figure}[t!]
\includegraphics[width=1.\columnwidth]{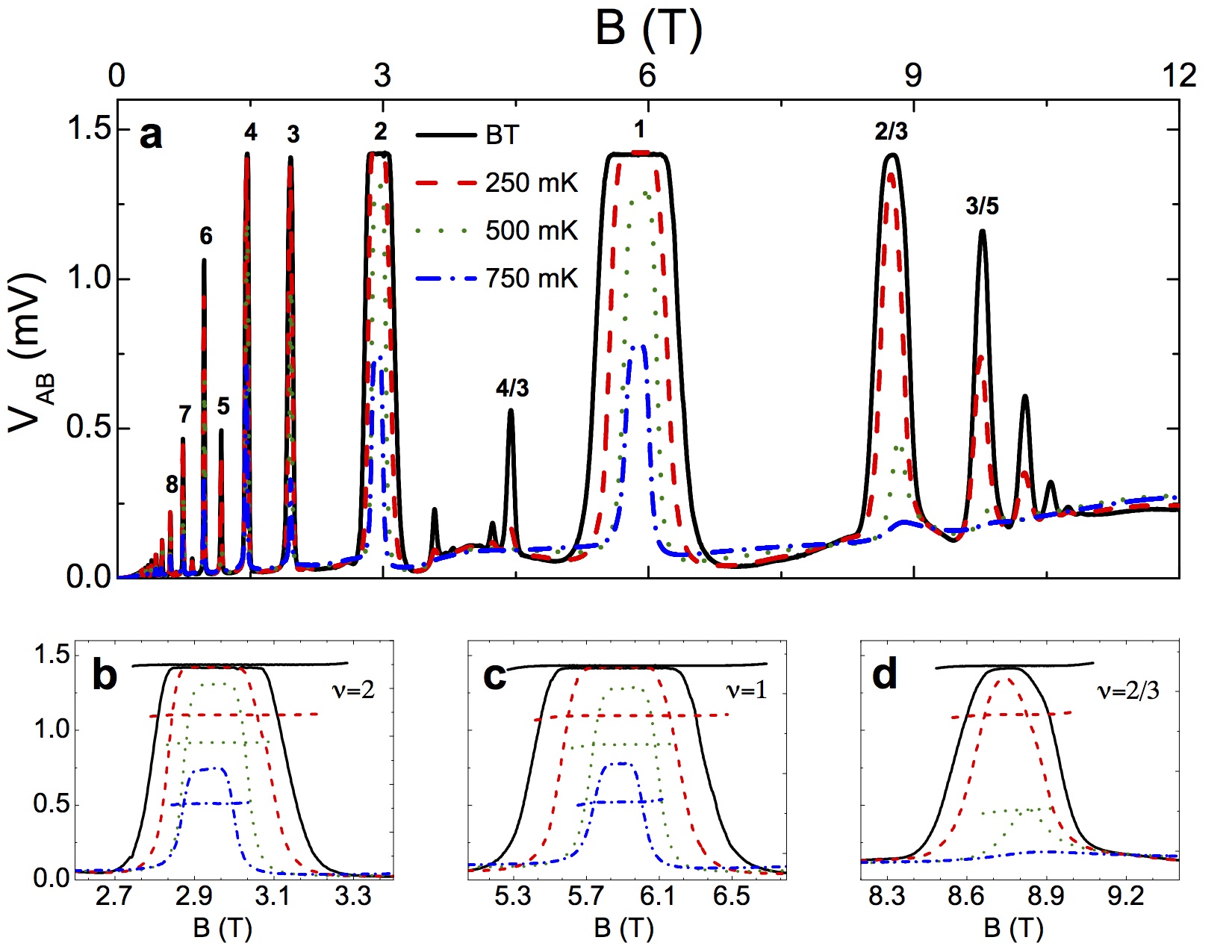}
\caption{\label{fig:3}\textbf{The temperature dependency of the inner voltage difference} \textbf{a,} The 12 Tesla wide span of $V_{\rm AB}$ for various temperatures, while excitations are $V_{\rm SD}=26$ mV and $V_{\rm in}=4$ mV fixed to sufficiently small voltages to prevent additional heating. \textbf{b-d,} The zoomed in filling factors together with the corresponding Hall plateaus.}
\end{figure}
In the following discussion we will assume spin degeneracy initially, since the mechanism to elucidate experiments is independent of the particular nature of the energy gap. We start our investigation with a situation where the lowest Landau level is partially occupied, i.e., the 2DES is completely compressible as depicted in Fig.~\ref{fig:2}f. The central panel in Fig.~\ref{fig:2} shows the measured potential difference between the inner contacts $V_{\rm AB}$ (broken (blue) line), the Hall potential $V_{\rm H}$ (solid thin line) and the potential difference between contacts 3 and 4 (thick (red) solid line), $V_{\rm L}$. In the graphical demonstrations (a-f) the electron density variation is depicted by the colour gradient (green), which vanishes (white) near both the inner (red square boxes) and the source-drain contacts (yellow, rectangular boxes). The black coloured areas correspond to incompressible (constant density) regions. It is worth to note that, sketching the density gradient and its behaviour near the contacts is well justified both experimentally and theoretically.~\cite{Dahlem:10,Aylin:unpub} We see that for $B\gtrsim3.4$ (case f) $V_{\rm H}$ and $V_{\rm L}$ increase (almost) linearly, whereas, $V_{\rm AB}$ considerably small. This behaviour suggests that the bulk of the 2DES acts as a poorly conducting metal, as predicted. We explain the behaviour of $V_{\rm AB}$ by modelling the bulk as serially connected resistances, composed of the inner contact resistance(s) $R_C$ and the resistance of the 2DES between contacts $R_B$. Decreasing the magnetic field, results in the quantisation of $V_{\rm H}$ and an increase in $V_{\rm AB}$, whereas $V_{\rm L}$ remains non-zero (case e). We elucidate these behaviours as follows: the region between contacts 1 and 4 becomes incompressible and decouples the Hall contacts and yielding a perfect quantisation. However due to the density gradient induced by the inner contacts, the entire bulk is not incompressible. We model this situation by a resistance and a capacitor ($C_{IB}$) connected in parallel. Since the capacitance approaches zero due to $D_T(E_F)=0$, all the measured potential results from $R_B$ and $R_C$. In addition there are regions between the contacts 3 and 4 where scattering takes place, yielding a non-zero $V_{\rm L}$. Notice that, $V_{\rm L}$ approaches zero in the e$_1$ case since the resistance along the leaky incompressible strip is reduced compared to a fully compressible bulk, as in case  e$_2$. Lowering $B$ further, results in formation of a bulk incompressible region, hence insulating state, spread all over the sample: case d. Here, QHE is well developed and $V_{\rm AB}$ becomes remarkably large and is bound by a cutoff voltage, which we attribute to finite TDOS at $E_F$. Namely, since $V_{\rm AB}=I.Z$, where $I$ is the excess current, the impedance reads $Z=\sqrt{R^2-(\frac{1}{2\pi i f e^2D_T(E_F)})^2}=$ constant. At a lower $B$ the bulk incompressible region splits into strips, two of which reside along the edges and the other two encircling the inner contacts. The edge strips decouple Hall contacts and, simultaneously, the encircling strips decouple the inner contacts due to their approximately infinite capacitance, resulting in a constant $V_{\rm AB}$, known as the Corbino effect.~\cite{Woltjer86:149} Note that, from the plotted measurements one cannot quantitatively determine the boundary between case (c) and (d), however, when we discuss the temperature effects this transition will become clear. The cardinal situation is where the encircling strips become thermodynamically compressible, i.e. similar to a leaky capacitor depicted by broken lines as depicted in case e$_1$, whereas the edge strips are still well decoupling as shown in Fig.~\ref{fig:2}b by solid strips.  In short, the evanescent encircling strips can be considered as leaky capacitors and the impedance between inner contacts is modelled similar to that of case (f). On the other hand, the edge strips are sufficiently wide, due to smoother edge density profile compared to that of in front of the inner contacts, and therefore decouple Hall contacts yielding a plateau. The edge strips connect contacts 3-4, hence, $V_{\rm L}$ is still zero. The situation in case (a) is trivial, where the edge strips also become transparent to Hall voltage and IQHE fades. Now, we are obliged to answer questions asked at the end of the previous Section. $V_{\rm AB}$ is constant only in a certain $B$ subinterval since the inner contacts are only decoupled by the incompressible bulk (case d) or by the encircling incompressible strips (case c) for a specific $B$ interval and the observed asymmetry is a direct result of incompressible state distribution depending on $B$. Namely at low field side it is due to encircling strips whereas at high field side it emanates from the bulk incompressible region. Our model also covers the $\nu=1$ and 2/3 plateaus, by lifting the assumption on spin effects and taking the energy gap to be the Zeeman gap for $\nu=1$ and the many-body gap for $\nu=2/3$. Interestingly, for all other visible plateaus we do not observe the saturated $V_{\rm AB}$, which suggests that the 2DES does not comprise an entire incompressible bulk. Note that, for $\nu=3$ and 4 the maximum value of the spikes hits the constant value $\sim$ 1.35 mV.

As the incompressibility conditions of the edge or encircling strips are critical to determine the transition from case (c) to (d), it is worth to investigate the effect of temperature on the observed features, by which we can also test our constant TDOS argument. One expects that, an increase in $T$ will result in decrease of the saturation value of $V_{\rm AB}$ which indirectly measures the available states at $E_F$. In addition, the incompressible regions will shrink in their widths at higher temperatures, hence the $B$ interval that we observe saturated $V_{\rm AB}$ will also decrease. This will allow us to determine qualitatively the transition between case (c) and (d). Fig.~\ref{fig:3} plots the temperature dependence of $V_{\rm AB}$ together with $V_{\rm H}$. The experimental findings strongly follow our expectations, such that the saturation value decreases ($\propto 1/D_T(E_F)$) meanwhile subinterval shrinks till the thermal energy overcomes the energy gap. Remarkably, we can identify the largest incompressible bulk at $B\sim2.9$ for $\nu=2$ from the maxima of $V_{\rm AB}$, where the slope traces the spatial extend of the bulk incompressible region. Notice that, the encircling strips decay faster due to thermal broadening of the TDOS, which makes easy for electrons to scatter across the incompressible strips, hence become leaky. Similar line of argumentation also holds for $\nu=1$, however, for the focused fractional state $\nu=2/3$ the maxima shifts to the high field edge of the plateau interval. We attribute this effect to different activation behaviour of the many-body gap. Temperature dependency of $V_{\rm AB}$ completes our experimental investigation.

\noindent\textbf{Conclusion}

\noindent We have reported on the bulk transport measurements of a geometrically modified Hall bar. On one hand we experimentally evidenced that it is possible to observe quantized Hall effect even if the bulk is not entirely incompressible, however, we simultaneously showed that it is also possible to observe quantized conductance if an incompressible strip resides at the edges preventing scattering between the Hall contacts. On the other hand, the incompressible bulk based topological theories are well justified at certain magnetic intervals also for finite size and non-periodically bound systems. In addition we showed that, the incompressible strips narrower than the thermodynamical length scales are prone to become leaky in the plateau-to-plateau transition intervals. For the IQHE, remarkably, there exists only a single incompressible strip with a given integer filling factor, in contradiction to single particle theories. Nevertheless, note that the Berry flux enclosed in real space comprised by the incompressible strip equals to the number of fully occupied Landau levels, hence, the Chern number is still determining the Hall conductance. New experiments based on the screening theory outlined here would be significantly important to determine interrelation between the topological insulators and quantized Hall effect which even could be extended to non-abelian states.

\section*{Supplementary Material}
In this supplementary material section we will first provide the details of our experimental setup and then discuss the validity of some crucial assumptions of the well known theories of the IQHE. Namely, first we will briefly discuss the assumptions on different boundary conditions and topological aspects of the quantized Hall samples within the single-particle (SP) theories. The main discussion is on the mapping between the momentum space representation of the edge states to real space representation. This discussion clarifies the importance of boundary conditions, where the normalisation condition is the only physical restriction. Here, instead of a detailed mathematical description we will make use of some schematic presentations which are common in the QHE discussions. However, to lift the confusion between different theories we used distinguishing colours to discriminate single-particle and screening theory concepts, e.g. incompressibility, edge states, etc. 

Our first discussion is based on the single-particle theory, where incompressible (namely energy gapped region or so to say insulating) bulk is essential for QHE and the incompressible regions are denoted by blue colour. We denote the SP chiral edge states by solid lines, which differ by colours depicting different filling factors, e.g. red corresponds to $\nu=2$ edge state where spin degree of freedom is neglected. On the other hand, consistent with the main text, we denote compressible 2DES by green and incompressible 2DES by black, within the screening theory. Yellow regions denote the contacts, whereas white corresponds to electron depleted (etched) regions for both theories. 

First, we aim to clarify that, using periodic boundary conditions for finite size Hall bars is questionable which then makes topological arguments prone to break. Second we would like to highlight that, considering quantum capacitance in calculating impedance or current modifies both the SP and screening theory results. And third, we explicitly show that our sample geometry is quite different then the well known Corbino and the anti-Hall bar geometries. 
\subsection{Experimental Setup}
In this paper our aim is to show that the bulk of the Hall bar is not incompressible through out quantum Hall plateaus by direct transport measurements. Our samples are defined on high purity GaAs/AlGaAs wafers, produced at Braun Center for Submicron Research. The Hall bars are 10 $\mu$ wide and 40 $\mu$m long, defined by chemical etching. Hall contacts are designed like a ``Lizard" such that they are not effected by the inner contacts. We embedded two inner contacts to the bulk of the Hall bar utilising air bridge technique, each inner contact has approximately 1 $\mu$m$^2$ area and they are 7.5 $\mu$m apart from each other. To measure Hall effect we excite the 2DEG by imposing a constant AC voltage between the source and drain contacts at  8.54 Hz by a Lock-in amplifier (SR 850, with an internal impedance of 10 M$\Omega$+25 pF). A 10 M$\Omega$ resistance is placed between inner contact A and source contact, whereas a 1 k$\Omega$ resistance is placed between inner contact B and drain contact. We measure Hall voltage (both X and Y components) between contacts 2-3 or 1-4. To check whether the Hall voltage is affected by inner contacts we also reversed source and drain contacts and observed similar features. Similar to Hall voltage measurements, we excite the inner contacts by a lock-in amplifier at 11.5 Hz, followed by a 10 M$\Omega$ resistance before contact A and 1 k$\Omega$ resistance after contact B. We measure the potential difference between the inner contacts, while a low-noise preamplifier is utilized to filter high frequency noise ($>$ 30 Hz). All the signals go through a room temperature RC circuit to filter noise above 92 kHz. We used 300 ms time constant and swept the magnetic field by a 0.1 T/min rate. The measurements are repeated for many thermal cycles and different samples defined on the same wafer. All the results were consistent.

\begin{figure}[t!]
\includegraphics[width=1.\columnwidth]{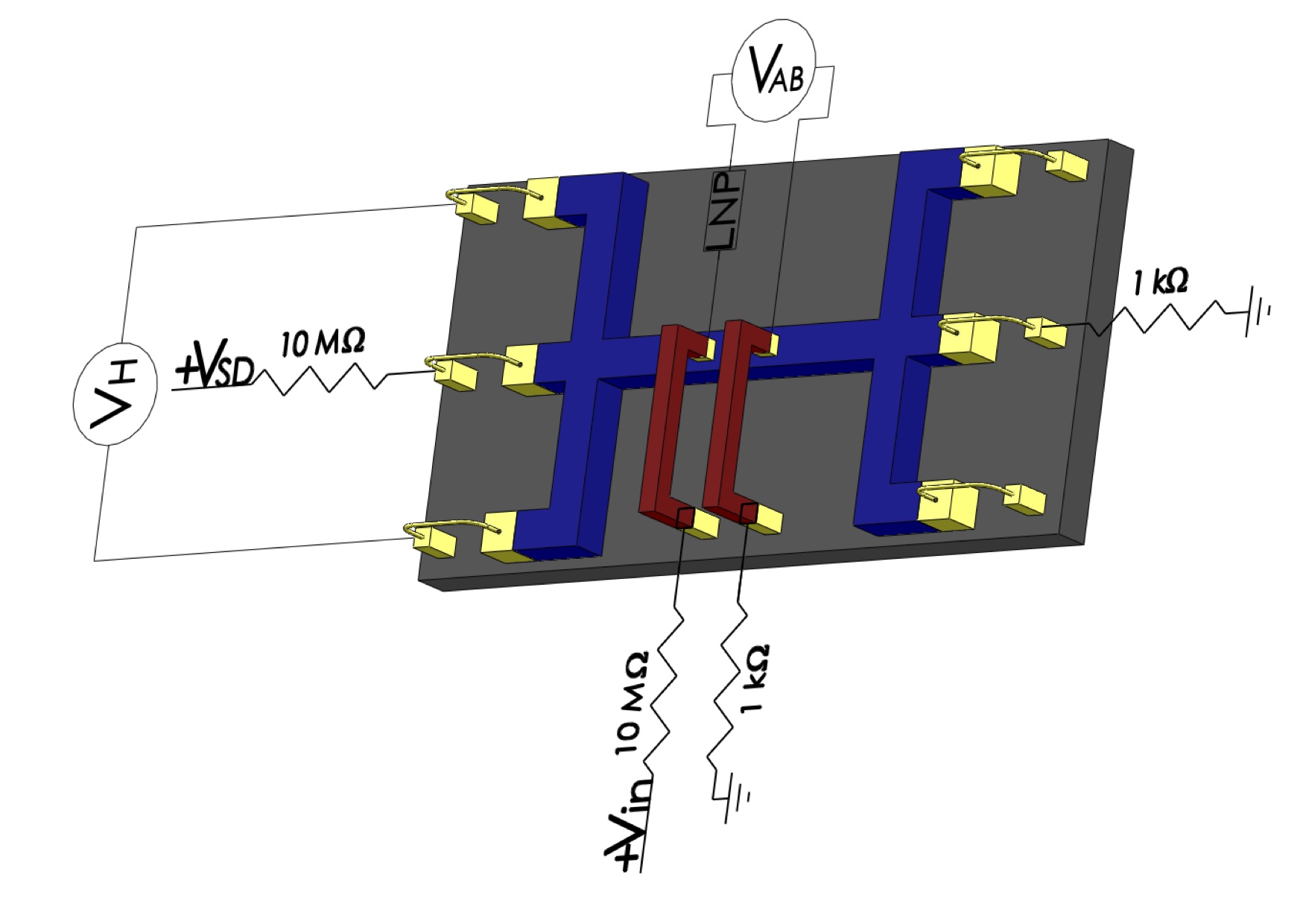}
\caption{\label{fig:add0}\textbf{Sample sketch.}  3D demonstration of our sample geometry, blue areas depict the 2DES, yellow regions are contacts and red denote the air bridges.}
\end{figure}

\subsection{Physical boundary conditions and their relation with Topology}
Once the Landau Hamiltonian is solved using Coulomb (also called Landau) gauge, i.e. translational invariance in $y$ direction and open boundary conditions in $x$ direction ($\Phi(x \rightarrow \pm \infty,y)=0$), the solution yields plane waves in the current ($y$) direction as if as a free electron and harmonic oscillator wave functions in the other direction. Such a choice of gauge and boundary conditions can be utilized to describe a homogeneous Hall bar that extends to infinity in both directions. However, to describe a more realistic Hall bar one usually assumes infinite walls in $x$ direction, which modifies the related wave-functions by parabolic cylinder functions approaching to simple harmonic oscillator solutions away from the boundaries. However, these preferences of boundary conditions yield the problem of normalisation. In order to overcome the normalisation problem one assumes periodicity in momentum along the current direction similar to Bloch wave function describing electrons in a crystal. As the Hall bar is not periodic in real space, one cannot impose periodicity in $y$, hence, periodicity in $k_y$ is assumed. This assumption yields to the well known description of Thouless~\cite{Thouless82:405} which explains quantised Hall effect in terms of Chern numbers in \emph{momentum} space, where one simply counts the Berry flux encircled and describes transport utilising the Kubo formalism. Such an approach is well justified only if the system is in electrostatic equilibrium, i.e. if no external current is imposed, or might be reasonable if the QHE can be treated within the linear response regime, if it can be handled within this regime at all. 
\begin{figure}[t!]
\includegraphics[width=1.\columnwidth]{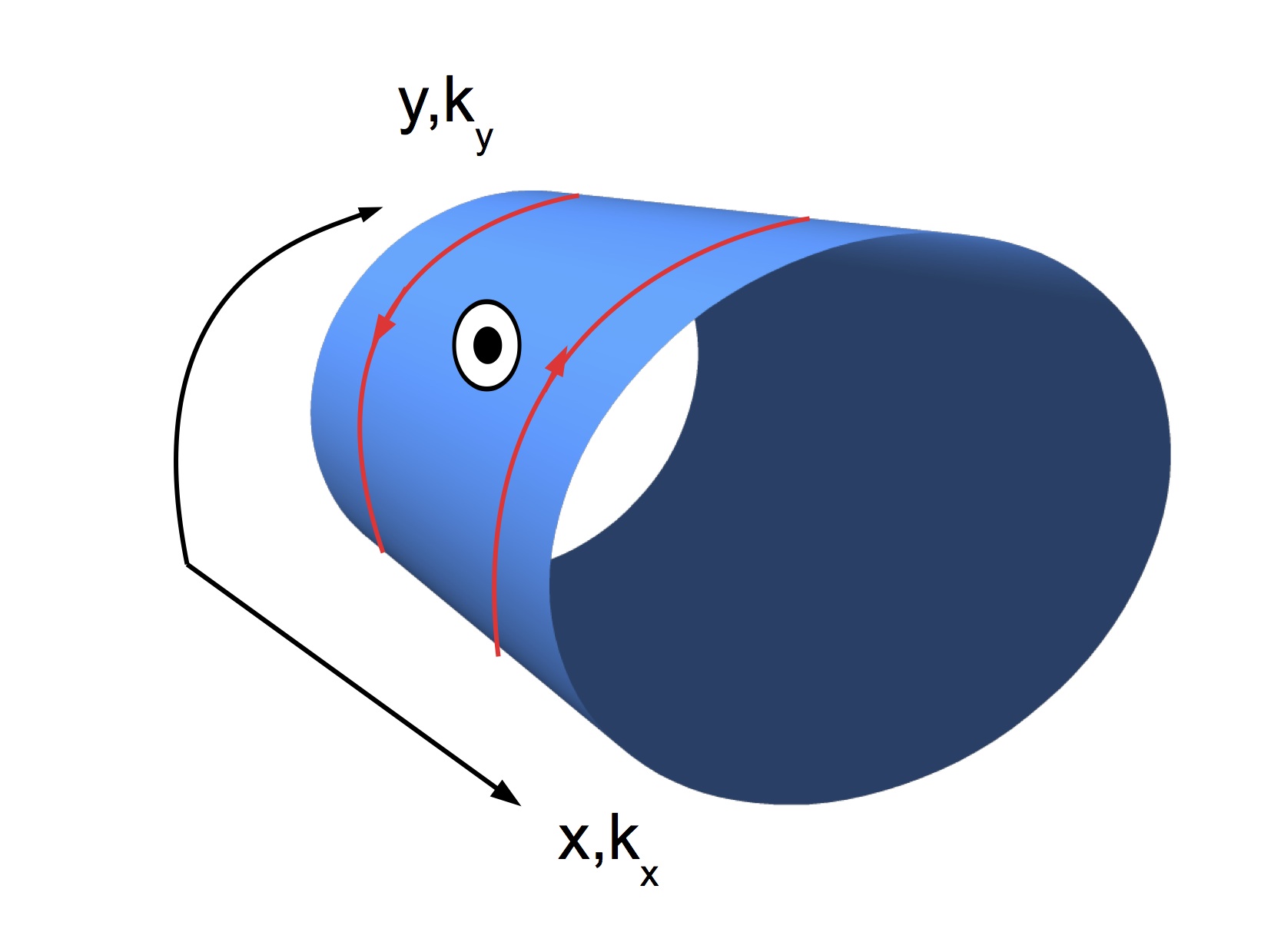}
\caption{\label{fig:add1}\textbf{Cylindrical geometry.} In the earliest attempt to elucidate IQHE, Laughlin proposed that if a 2DES resides on the surface of a cylinder and a time dependent perpendicular magnetic field penetrates the surface, then to protect the geometrical phase only an elementary charge can be moved from one end of the cylinder to the other end when the flux changes by one quantum. In this description periodicity in real space is assumed, however, the boundary conditions are not specified in the perpendicular direction. In contrast, in the approach of Thouless the periodicity is imposed in momentum space $k_y$, where the 2DES is defined on a flat surface.}
\end{figure}
The first case can be ruled out in our experiments since we excite the system by an external voltage. The second case is also ruled out, since the excitation energy (voltage) is comparable with the Landau gap, hence the system is far away from linear response. In contrary to periodicity in momentum space, Laughlin assumed periodicity in real space considering a cylinder where instead of an imposed current he assumed a radial magnetic field which changes adiabatically by time (cf. Fig.~\ref{fig:add1}), leading to a phase which is protected by the topology resulting in conductance quantisation.~\cite{Laughlin81} This approach also for sure does not entirely describe the experimental realisation of the QHE, at least for our experiments. Both of the descriptions of the QHE implicitly require a bulk incompressibility, namely an energy gap induced by the magnetic field. This is somewhat similar to 3+1 D topological insulators where an energy gap opens due to crystal structure and provides an incompressible (insulating) bulk.~\cite{Hasan:10:3045} However, note that this gap is in the energy dispersion, namely when one plots the relation between the energy and momentum there opens a gap due to symmetries of the crystal. Similarly, for an infinite Hall bar with periodic boundary conditions one can obtain such a gapped energy dispersion in momentum space, which is usually mapped to real space without taking care of the boundary conditions of the physical devices. For a finite width and length (flat) Hall bar assuming periodic boundary conditions is not appropriate.  Hence, bulk incompressibility which is essential to describe QHE does not apply to our physical system at hand. In connection with 3+1 D topological insulators, we should note that in this case the system is periodic in (crystal) momentum space and hence the energy dispersion presents the well appreciated ``topologically" protected energy gap. In contrast, QHE does not provide the periodicity in momentum space once realistic, i.e. physical, boundary conditions are considered.

\subsection{Real space-momentum space duality of Halperin-B\"uttiker edge states}
\begin{figure}[t!]
\includegraphics[width=1.\columnwidth]{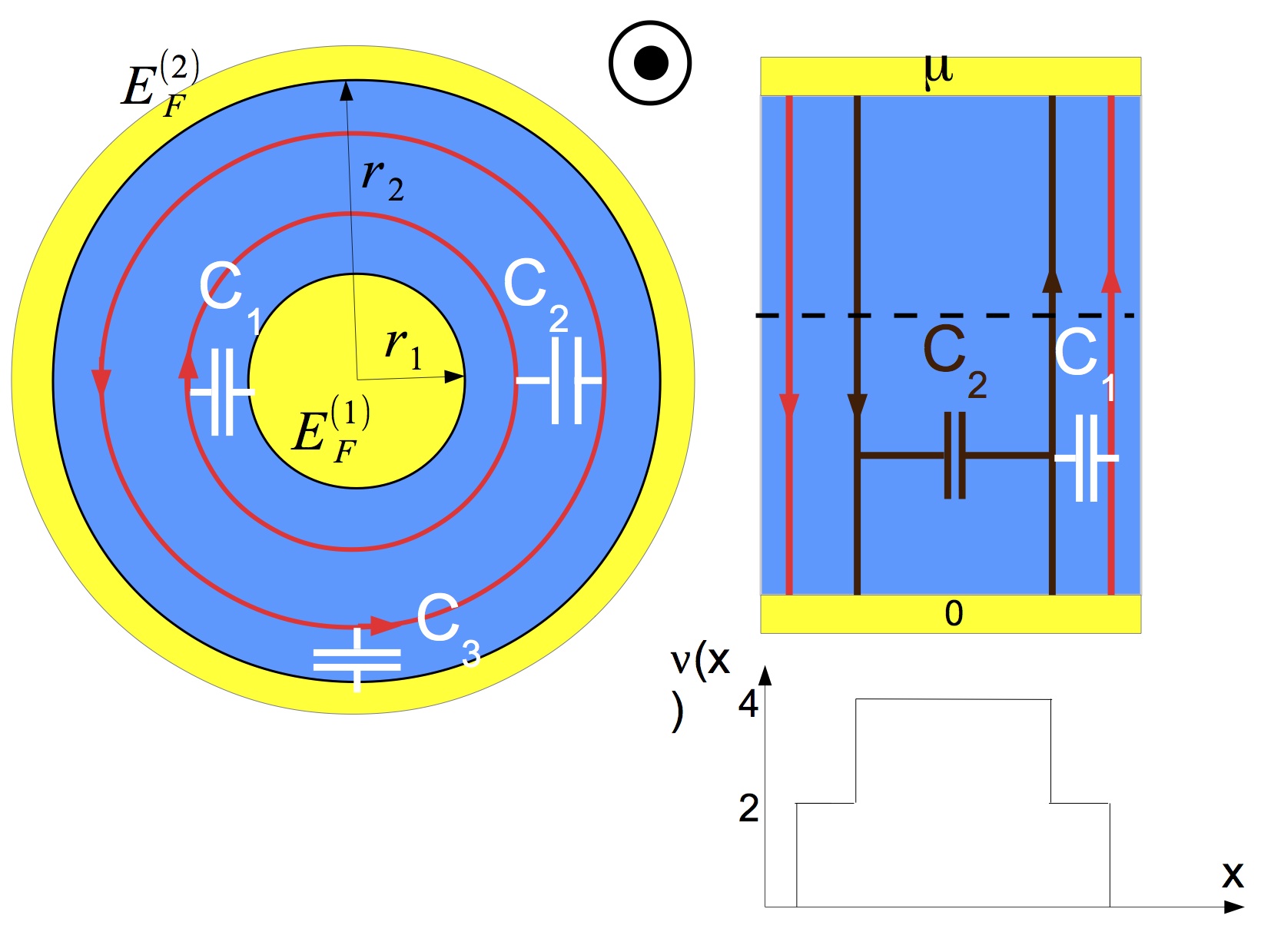}
\caption{\label{fig:add2}\textbf{The Corbino and Hall bar geometries.} The quantised Hall effect measurements are usually performed either on the rotationally periodic Corbino \textbf{(a)} geometry or on the translationally non-invariant Hall bar \textbf{(b)} geometry in real space. For a Corbino disc, inner and outer contacts (yellow regions) are kept at different electrochemical potentials, where edge states (red solid lines) are parallel to contacts and the conductance between these contacts is measured. In contrary, for a Hall bar geometry edge states are perpendicular to the contacts, where contacts are kept at different electrochemical potentials, here 0 and $\mu$. The bottom panel in \textbf{(b)} presents the filling factor (or electron density) distribution along the cut denoted by dashed lines in the upper panel. One clearly sees the unrealistic electron distribution which for sure cannot guarantee electrostatic equilibrium. Note that, the electron depleted regions near the edges are neglected, hence the related capacitances, which is consistent with the single-particle theories.}
\end{figure}
The first important manifestation of boundary conditions on the transport properties of a 2DES subject to high magnetic fields is proposed by Halperin in early 80's.~\cite{Halperin82:2185} In his pioneering work he imposed the above mentioned infinite wall boundary conditions in the radial direction $r$, which essentially bends the Landau levels in the close vicinity of boundaries. Utilising periodicity in azimuthal ($\theta$) direction (both in momentum and real space), one can map the energy dispersion (i.e. energy versus momentum) to real space (i.e. energy versus $r$) and obtain the geometry known as the Corbino geometry, as shown in Fig.~\ref{fig:add2}a. In the original work it is stated that this is a ``slight modification" of Laughlin's cylindrical geometry (see Fig.~\ref{fig:add1}), which is not reasonable from topological point of view in real space, since the genus numbers are different (for Laughlin's case it is 1, since periodicity is also assumed in $z$ which defines essentially a torus, and for Halperin's case it is 0). We should also note that, due to the absence of edges in Laughlin's geometry, the current is uni-directional, whereas at a Corbino edge states are present and carry chiral current. For a Corbino geometry, in contrast to a Hall bar (cf. Fig.~\ref{fig:add2}b), there are no source and drain contacts, however, the edge states exist due to boundaries where all the system is in equilibrium. Namely, no external current can be and is imposed. In this geometry, one can measure the conductance between the inner and outer contacts if contacts are kept at different electrochemical potentials. In such a system the electrical measurements essentially results in a quantised Hall conductance. It is also stated that ``In a real experiment, the measured Hall potential $eV$ is the sum of an electrostatic potential $eV_0$ and the difference in Fermi levels $E_F^{(2)}-E_F^{(1)}$ (electrochemical potential difference in our notation). The edge current is then only a \emph{fraction} of the total Hall current, given by \be (E_F^{(2)}-E_F^{(1)})/eV \approx \alpha n r_c\hbar\omega_cC/e^2,\label{eq:halperin}\ee where $C$ is the capacitance per unit length of the edge states, and $\alpha$ is a number of unity.". Here, $n$ is an integer determined by the filling factor $\nu$ at the bulk and Fermi energy lies in between the energies $E_{\nu}$ of two Landau levels $\nu=n-1$ and $\nu=n$, in the interior of the sample. The electrochemical potentials near the boundaries ($E_F^{(2)}$ and $E_F^{(1)}$, at $r_2$ and $r_1$, respectively, where the contacts reside) are also supposed to lie in the interval $E_{n-1}-E_n$, which implicitly assumes linear response. Then the total current carried by the edge states between $E_F^{(2)}$ and $E_F^{(1)}$ is given by $neh^{-1}(E_F^{(2)}-E_F^{(1)})$. Now, we should clarify couple of points with the properties of contacts and their equilibration with edge states: first, in Halperin's approach it is implicitly assumed that the capacitances $C_1$ and $C_3$ shown in Fig.~\ref{fig:add2}a are neglected, which imposes that the contacts are in electrochemical equilibrium with the edge states, namely the scattering between the edge states and contacts is possible pointing a compressible edge. Hence, the contribution to Hall current in real experiments only come from the bulk capacitance, which is quantised if finite TDOS at $E_F$ is assumed. If the compressible edge assumption is lifted, then the total capacitance would be the sum of $C_1$, $C_2$ and $C_3$. For $\nu_{centre}=2$ the total capacitance would yield an edge current approximately $\frac{2}{3}\alpha r_c\hbar\omega_cC/e^2$, which is not quantised. Here we assumed $D_T(E_F^{(1)})=D_T(E_F^{(2)})=D_T(E_F)$. For higher bulk filling factors, such an approach implicitly assumes that the inner edge states are in equilibrium with each other similar to the outer edge states, which again requires that the edges are compressible, i.e. scattering between inner or outer edge states is possible. This assumption then makes the current quantisation questionable. 

As discussed in the main text and details given in the following Subsection, capacitance is composed of the classical and the quantum counterparts. Once, the region between two edge states (or contacts) is insulating (namely incompressible) then the quantum capacitance vanishes for an ideally pure system. Hence the Hall current mentioned above reads to zero, i.e. impedance diverges. In real experimental devices there are potential fluctuations due to disorder which yield localised states as discussed by Halperin, which then results in conductance quantisation for a finite magnetic field interval, where bulk incompressibility is still preserved. Furthermore, once again the main assumption is the incompressibility of the bulk, i.e. the impedance between the inner and the outer contacts reads to infinity (or relatively high compared to conductance) hence the conductance measured between these contacts should be quantised. To be explicit, if one wants to move an electron from one contact to the other one has to get across $n$ edge states (namely pass through metal-topological insulator boundary $n$ times) and has to pay an amount of energy that corresponds to $n$ (Berry) flux quanta, yielding to quantisation. In Fig.~\ref{fig:add2} we present both the Hall bar and the Corbino geometries also depicting ``edge states". Note that, in the Corbino geometry edge states are parallel to the contacts whereas for the Hall bar geometry they are perpendicular to the contacts. Hence, the impedance is infinite (or huge for real devices at finite $T$) for the Corbino geometry, however, for the Hall bar capacitance vanishes (or much smaller then the resistance) considering a ``B\"uttiker" contact, i.e. transmission from the contact to edge state is unity.

It is important to note that, the mathematical mapping between the momentum and real space representations of the edge states makes sense only if the periodicity is preserved, which is the case for the Corbino geometry. However, when a finite size Hall bar is considered such a mapping becomes questionable, at least for all geometries. 

Couple of years later than the work of Halperin, B\"uttiker developed a transport theory based on the Landauer formalism to describe the IQHE.~\cite{Buettiker86:1761} In his work also the finite size of the Hall bar is taken into account by imposing a confinement potential in $x$ direction, which essentially varies smoothly on quantum mechanical length scales. Although yielding to similar results with Halperin, the smooth confinement potential allows one to use simple harmonic oscillator solutions also close to the boundaries. In addition, the scattering probability between edge states at the same boundary is suppressed by the fact that the edge states are relatively far apart from each other both in real space and in momentum space. Utilising the periodicity in momentum space in equilibrium one can then map the momentum space representation of edge states to real space and draw the well known ``B\"uttiker" edge states for a Hall bar as shown in Fig.~\ref{fig:add2}b. This picture is modified if an external current is imposed by applying a voltage difference $\Delta V$ between the source and drain contacts, yielding an electrochemical potential energy difference $\mu=eV$. Of course in this non-equilibrium situation assuming periodic boundary conditions become questionable, where one end of the Hall bar is kept at potential $V$. Turning back to our discussion on the capacitances and calculating the actual Hall current one can still utilise Eq.~\ref{eq:halperin}. Note that the edge states at the same sides have the same electrochemical potential, hence, $C_1$ and $C_3$ vanishes and the only contribution to capacitance comes from $C_2$. Once again, for an ideal 2DES capacitance vanishes yielding zero Hall current pointing a back-scattering free transport throughout the plateau interval. However, even a small amount of TDOS below $E_F$ would yield a finite capacitance, hence a finite Hall current which would result in deviations from perfect quantisation. 

By the above discussion, we have shown that the momentum-real space duality of both Halperin and B\"uttiker edge states becomes questionable since such a duality strongly depends on the imposed boundary conditions and the symmetries of the sample. In addition, once the quantum capacitance is taken into account, which is directly proportional to TDOS at $E_F$, impedance between probe contacts also strongly depends on the edge state configurations, namely whether the edge states are perpendicular or parallel to imposed current. Such geometrical and topological aspects of so called QHE samples are usually undermentioned in the well known theories.    
\subsection{Classical and Quantum capacitances between contacts}
\begin{figure}[t!]
\includegraphics[width=1.\columnwidth]{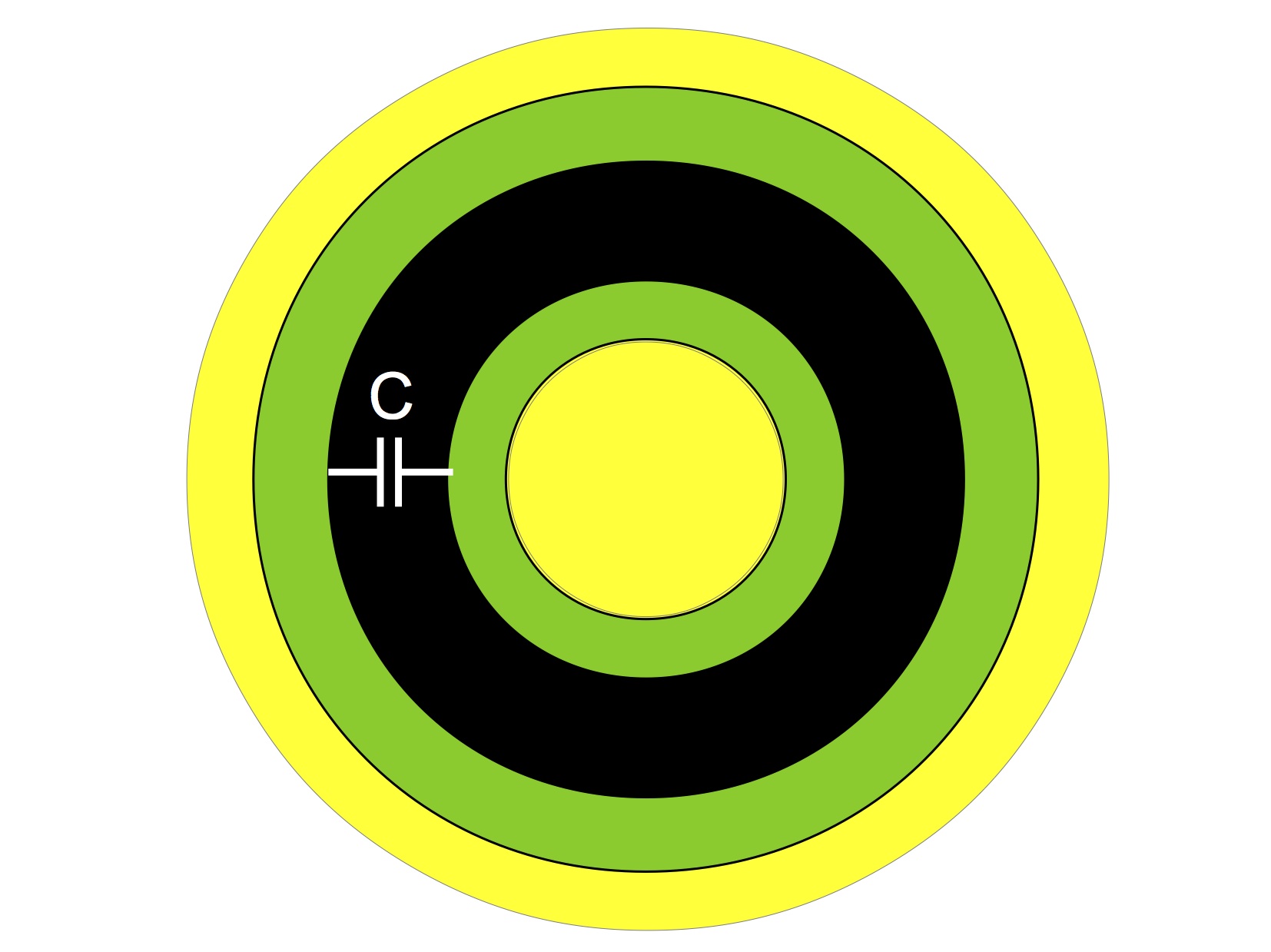}
\caption{\label{fig:add3}\textbf{Screening theory expectation of a Corbino disc.} The inner and outer contacts are decoupled by a bulk incompressible region denoted by black. However, if the magnetic field is reduced the picture essentially does not change drastically. At a lower $B$ field, the bulk incompressible region will split into two encircling incompressible rings near the edges and since the capacitances are connected in series the contacts would be still decoupled.}
\end{figure}
In a simplistic treatment, inverse capacitance can be described as the amount of energy to add a charge to a system, given by $Q^2/C_c=E$. Classically, one can approximate the capacitance of a sheet as shown in Fig.~\ref{fig:add3} as $C_c=\frac{\epsilon L}{2\pi^2}\ln{\frac{4d}{w}}$, where $\epsilon$ is the dielectric constant, $L$ is the perimeter of the ring, $w$ is the width of the incompressible strip and $d$ is the distance between the gate and the 2DES.~\cite{Evans:93} On the other hand, the quantum capacitance per area is solely dependent on the thermodynamical density of states given by $C_q=e^2D_T(E_F)$. Since these two capacitances are connected in series, the total capacitance is given by $1/C=1/C_c+1/C_q$. The classical capacitance is finite except that the width of the incompressible region is infinite, however, the quantum capacitance becomes zero if there exists an incompressible strip between the inner and outer contacts at zero temperature and for an ideally clean system, i.e. no level broadening due to impurities. Therefore, the capacitance is dominated by the quantum counterpart if there exists an incompressible strip decoupling the contacts. Note that, the contacts and the compressible regions are in electrochemical equilibrium, since they both behave like a metal. Furthermore, since scattering is suppressed exponentially along the incompressible ring and is quantised across the ring, resistance is much smaller than the total capacitance therefore impedance reads 
\be Z \propto 1/D_T(E_F) \rightarrow \infty . \ee
Hence, for a Corbino geometry the bulk incompressibility is the guarantee of conductance quantisation within the single particle picture. However, when interactions are taken into account we observe that it is not necessary to have an incompressible bulk and only a single incompressible strip is sufficient to decouple inner and outer contacts.

The next discussion is on the thermodynamical definition of incompressibility which is a statistical quantity. One usually defines a system to incompressible if the ratio between the change in number of particles and the change in electrochemical potential presents a discontinuity. Note that, electrochemical potential is a statistical quantity which makes sense physically only if there are sufficient number of particles within the system considered. Therefore, assigning incompressibility to a strip is bounded by the number of particles within the strip, which is essentially determined by the Fermi wavelength at zero temperature. Hence, once the strip width becomes small or comparable with the Fermi wave length the insulating behaviour of the strip is destroyed. Such a situation can be considered as a leaky capacitor, as stated in the main text. In addition, when the strip width becomes even smaller than the magnetic length, which is the quantum mechanical length scale, then it is possible to tunnel across the incompressible strip by quasi-inelastic scattering mechanisms.   
\subsection{What to expect within the B\"uttiker edge state picture between inner contacts}
\begin{figure}[t!]
\includegraphics[width=1.\columnwidth]{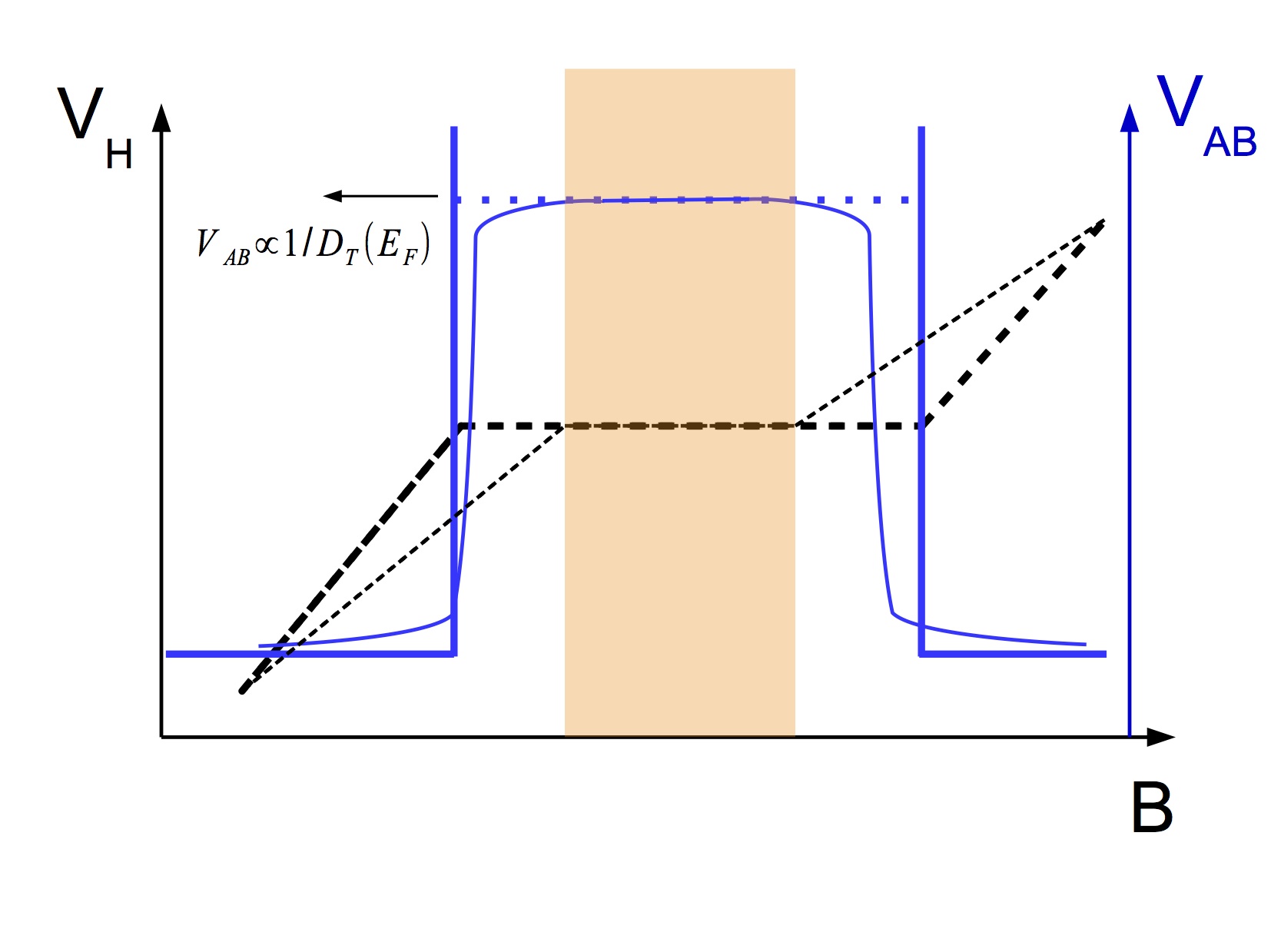}
\caption{\label{fig:add4}\textbf{The relation between $V_{\rm H}$ and $V_{\rm AB}$ as a function of $B$ field.} Based on the above discussion concerning capacitances one expects to observe an infinite $V_{\rm AB}$ whenever the Hall voltage is constant since the bulk should be incompressible. Due to $D_T(E_F)\neq0$ at realistic samples the impedance becomes finite. In addition, at finite temperatures the sharp features both at  $V_{\rm H}$ and $V_{\rm AB}$ are smeared out. However, the constant Hall voltage and high impedance intervals should still overlap.}
\end{figure}
\begin{figure}[t!]
\includegraphics[width=1.\columnwidth]{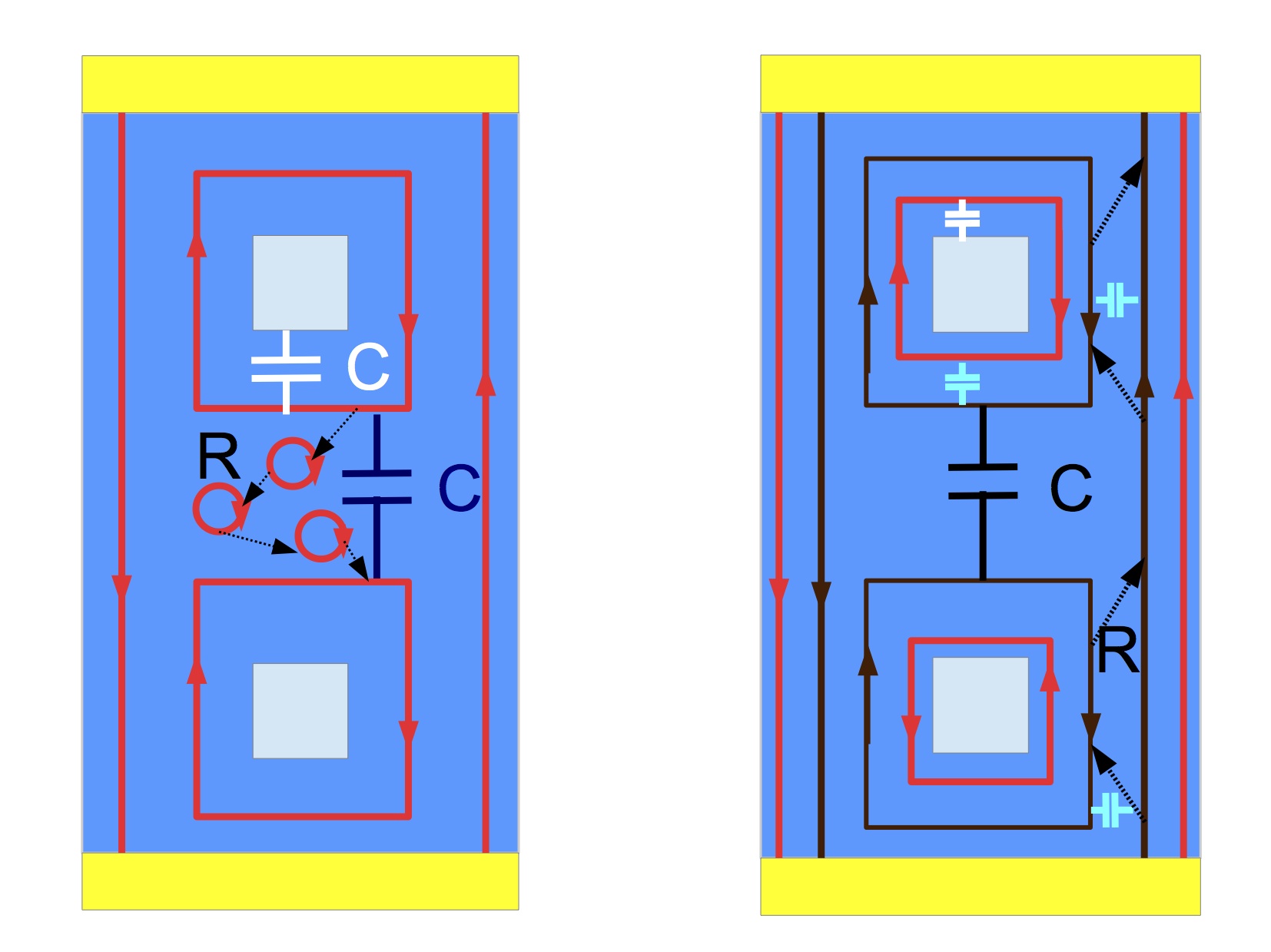}
\caption{\label{fig:add5}\textbf{The edge state distributions and capacitances within the SP theories.} Sketches present edge states considering different average filling factors, \textbf{(a)} $\nu=2$ and \textbf{(b)} $\nu=4$, where also the capacitances between the edge states and contacts are presented by inner (white) symbols. Similar to Halperin's discussion of a Corbino geometry, the capacitance between contacts are determined by the bulk incompressible region, however, here the excitation is perpendicularly imposed with respect to edge states. Interestingly, even at the transition between $\nu=4$ to $\nu=2$ plateau the inner edge states should provide a high capacitance. Namely one can expect scattering between $\nu=4$ edge states, which would yield a huge impedance even at the transition intervals. }
\end{figure}

As discussed above, once the QHE is well developed namely scattering is suppressed by the incompressible bulk the impedance between contacts should diverge for an ideally pure 2DES. Hence, if Hall voltage presents quantisation $V_{\rm AB}$ should diverge simultaneously. However, this is true only at zero temperature and for an ideally pure 2DES. Once, effects of finite temperature and TDOS at the Fermi energy is considered then the Hall plateau shrinks together with the high impedance interval. Let us first consider $T=0$ however assume a non-ideal 2DES, i.e. with impurities leading to level broadening, then the amplitude of the impedance (therefore $V_{\rm AB}$) is bounded from up by the TDOS at $E_F$. Fig.~\ref{fig:add4} depicts the upper boundary of $V_{\rm AB}$ at zero temperature by the thick horizontal dotted line, whereas ideally pure 2DES case is shown by the solid (blue) line. Once the impurity scattering is taken into account at finite temperature the curves are modified, which are depicted by the thin broken lines for $V_{\rm H}$ and by the thin solid lines for $V_{\rm AB}$. Based on above arguments one can conclude that our experimental findings do not coincide with the results expected from B\"uttiker type edge state theory, up to our knowledge. We show the expected edge state distribution and corresponding capacitances in Fig.~\ref{fig:add5}. The scattering between edge states out of the plateau interval is depicted by $R$ denoting the resistance. Here one can see once more the geometrical difference between the Corbino geometry (one hole) and our device (two holes). In the latter both contacts are encircled by the edge states where excess current is perpendicular to the contacts, whereas in the former one the outer edge state is encircled by the outer contact and there is no current imposed between the two contacts. The quality of our sample guarantees that the electrochemical equilibration between the edge states is strongly suppressed given the fact that the distance between inner contacts is 3.5 $\mu$m, the area of contacts being 1 $\mu$m$^2$ and the physical dimensions of the our device is ($W\times L$) $10\times50$ $\mu$m$^2$. In our sample the mean free path, the localisation and the equilibration lengths are sufficiently large to suppress scattering between edge states both encircling the contacts and the ones along the edges of the sample. 
\subsection{Differences between Anti-Hall bar and our geometry}
\begin{figure}[t!]
\includegraphics[width=1.\columnwidth]{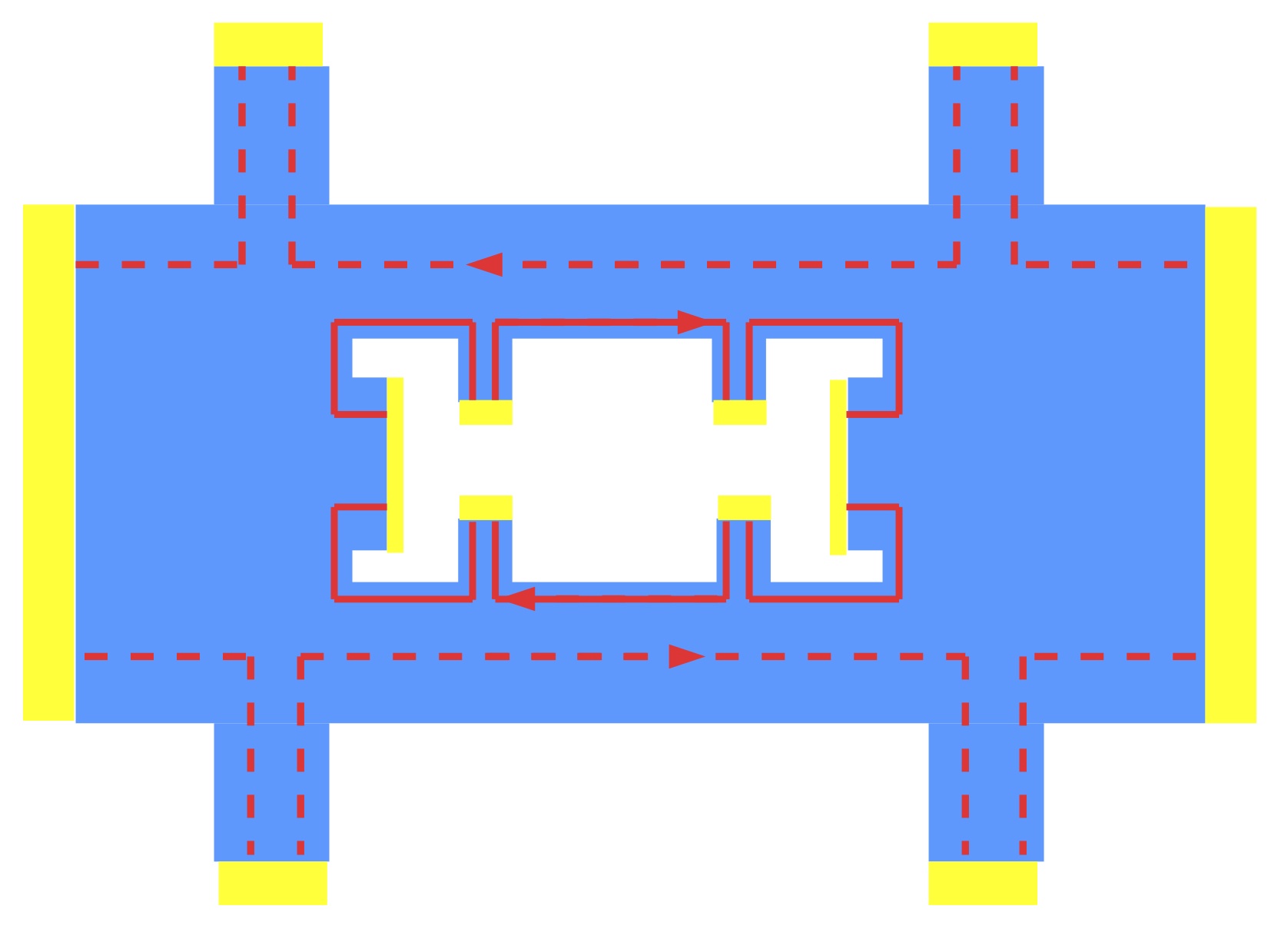}
\caption{\label{fig:add6}\textbf{Sketch of an anti-Hall bar.} Once an anti-Hall bar is ``embedded" into a Hall bar, one expects to have two independent edge state configurations which are decoupled by the incompressible region in between. Here it is essential to realise that, the edge states of the anti-Hall bar connect the source and drain contacts and do not encircle entirely any of the contacts.}
\end{figure}
Modifying the topology of a Hall bar has been realised some 18 years ago by R. Mani~\cite{Mani:96} and recently modelled by Oswald and his co-workers within the non-equilibrium network model.~\cite{Oswald:12} At a first glance our experimental system and the so called anti-Hall bar (AHB) geometry seems to resemble each other, as shown in Fig.~\ref{fig:add6}. However, there are couple of major differences, first at an AHB geometry there is a single etched region at the bulk of the sample. In this configuration edge states percolate around the AHB and contacts are connected by the edge states. However, in our situation we have two electron depleted regions, i.e. inner contacts, which are not connected by incompressible strips. Hence for the AHB geometry, impedance is mainly dominated by the resistive counterpart, whereas for our geometry the transport between contacts is dominated by the capacitive counterpart throughout the plateau interval. Another important difference is the properties of our sample, namely its size and the mobility. Different from the actual AHB sample, our samples are defined on a high purity wafers which strongly suppress scattering between edge states, either between the inner ones or between the inner and outer ones.  
\subsection{Finite density of states below the Fermi energy, indirect proof of a single incompressible strip and the 1.35 mV value}
\begin{figure}[t!]
\includegraphics[width=1.\columnwidth]{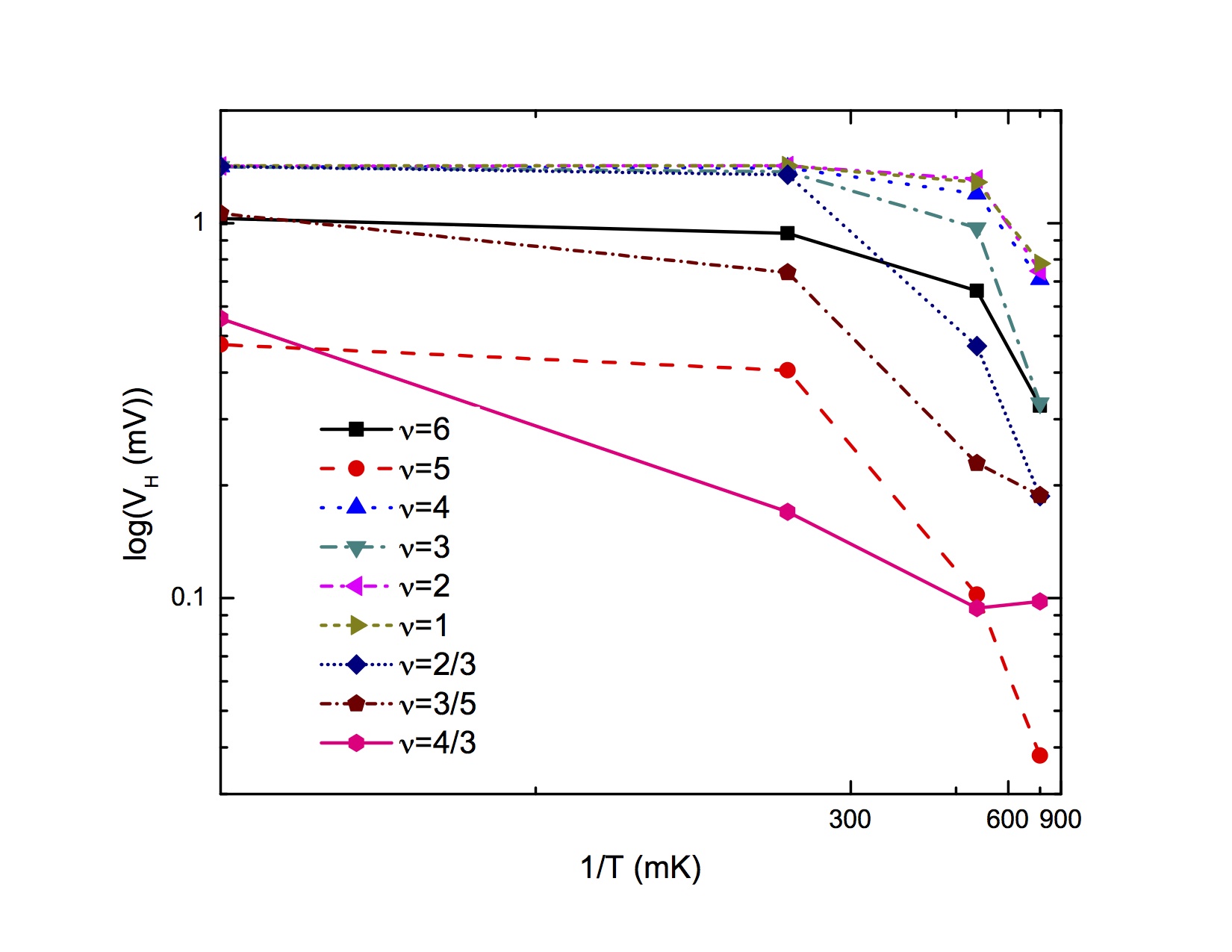}
\caption{\label{fig:add7}\textbf{The Arrhenius plot of Max($V_{\rm AB}$).} The maximum of the voltage drop between contacts A and B considering different filling factors. One can see that, the activated behaviour of the well developed integer states coincide nicely, whereas for fractional states the gap value is scattered depending on the filling factor.}
\end{figure}
From the above discussions one can clearly see that the potential difference measured between contacts A and B, should diverge or at least become huge at certain $B$ intervals within the quantised Hall plateau. Consequently, the maximum value of $V_{\rm AB}$ in certain magnetic field intervals is attributed to finite TDOS at the Fermi energy. We also observe that, this maximum value is not bulk filling factor dependent and is not quantised. The non-quantisation is clarified by the temperature dependent measurements. At base temperature and at the lowest inner excitation voltage ($V_{\rm in}=4$ mV), we observe that Max($V_{\rm AB}$)$\approx1.35$ mV for the plateaus $\nu=4,3,2,1$ and 2/3, whereas it seems that for the other quantised Hall states only a spike is measured. Now let us consider a situation where we imposed an excitation at very low frequencies (practically DC) and let us assume that the an infinite resistor (impedance) is placed between the inner contacts at zero temperature, then we would expect to measure 2 mV, i.e. one has basically 2 to 1 voltage divider. Hence, the maximum value indicates that the bulk is not an infinite resistor, however, it highly decouples inner contacts due to finite TDOS at $E_F$. In the next experimental investigation we measured $V_{\rm AB}$ as a function of temperature and observed that Max($V_{\rm AB}$) decreases by increasing $T$. The decrease of the maximum value is a direct consequence of finite TDOS at $E_F$ and is a measure of the activation gap, indirectly. In Fig.~\ref{fig:add7}, we show the Arrhenius plot of Max($V_{\rm AB}$) and see the commonly observed activated behaviour. We also checked the value of Max($V_{\rm AB}$) at higher excitation voltages and see the same ratio between the Max($V_{\rm AB}/V_{\rm in}$), indicating that this maximum value is not due to a cutoff voltage of our electronic setup. Both of these experimental results clarify that the bulk resistance (in fact voltage difference at fixed excitation current) is not quantised. One can also claim that there exists only a single incompressible encircling strip with a given (integer or fractional) filling factor, since if there existed more than one strip Max($V_{\rm AB}$) would present a stepwise behaviour at each integer filling factor plateau.   

\acknowledgments
We would like to thank J. Jain and K. von Klitzing for initiating the idea of embedding inner contacts. A.S. also acknowledges ITAP-Marmaris for providing a perfect scientific environment, where the final version of the paper was edited.
This work is supported by T\"UB\.ITAK under grant (112T248 and 211T164) and Istanbul University scientific projects department IU-BAP:6970 and 22662.
\bibliographystyle{ieeetr}

\end{document}